\documentclass[conference]{IEEEtran}
\IEEEoverridecommandlockouts
\usepackage{ifpdf}
\usepackage[dvips]{graphicx}
\graphicspath{{../eps/}}
\DeclareGraphicsExtensions{.eps}

\usepackage{mathbbol}
\usepackage{cite}
\usepackage[T1]{fontenc} % optional
\usepackage{algorithm}
\usepackage{algorithmic}
\usepackage[cmex10]{amsmath}
\usepackage{amsthm}
\usepackage{amssymb}
\usepackage{balance}
\usepackage{eqparbox}
\usepackage{textcomp}
\usepackage{amsfonts}
\usepackage[eulergreek]{sansmath}
\usepackage{color}
\usepackage[cmintegrals]{newtxmath}

\newtheorem{theorem}{Theorem}
\newtheorem{remark}{Remark}
\usepackage[open,openlevel=1]{bookmark}
\usepackage{enumitem}
\usepackage{scalerel}

\allowdisplaybreaks

\newcommand{\mb}[1]{\mathbf{#1}}

\makeatletter
\newcommand\fs@betterruled{%
	\def\@fs@cfont{\bfseries}\let\@fs@capt\floatc@ruled
	\def\@fs@pre{\vspace*{6pt}\hrule height.8pt depth0pt \kern2pt}%
	\def\@fs@post{\kern2pt\hrule\relax}%
	\def\@fs@mid{\kern2pt\hrule\kern2pt}%
	\let\@fs@iftopcapt\iftrue}
\floatstyle{betterruled}
\restylefloat{algorithm}
\makeatother

\newcommand{\bseq}{\begin{subequations}}
	\newcommand{\eseq}{\end{subequations}}
\newcommand{\baln}{\begin{align}}
	\newcommand{\ealn}{\end{align}}
\newcommand{\balnd}{\begin{aligned}}
	\newcommand{\ealnd}{\end{aligned}}
\newcommand{\beq}{\begin{equation}}
	\newcommand{\eeq}{\end{equation}}
\newcommand{\beqn}{\begin{eqnarray}}
	\newcommand{\eeqn}{\end{eqnarray}}
\newcommand{\beqno}{\begin{eqnarray*}}
	\newcommand{\eeqno}{\end{eqnarray*}}
\newcommand{\bma}{\begin{displaymath}}
	\newcommand{\ema}{\end{displaymath}}
\newcommand{\bnu}{\begin{enumerate}}
	\newcommand{\enu}{\end{enumerate}}
\newcommand{\bce}{\begin{center}}
	\newcommand{\ece}{\end{center}}
\newcommand{\btb}{\begin{tabular}}
	\newcommand{\etb}{\end{tabular}}
\newcommand{\ba}{\begin{array}}
	\newcommand{\ea}{\end{array}}

\begin{document}
\bstctlcite{IEEEexample:BSTcontrol}
\title{Large-Scale Beam Placement and Resource Allocation Design for MEO-Constellation SATCOM} 
\author{{\vspace{-8mm}}\\
	\IEEEauthorblockN{Vu Nguyen Ha${}^{\dagger}$, Eva Lagunas${}^{\dagger}$, Tedros Salih Abdu${}^{\dagger}$, Haythem Chaker${}^{\dagger}$, Symeon Chatzinotas${}^{\dagger}$, Joel Grotz${}^{\ddagger}$} %\\
\IEEEauthorblockA{\textit{${}^{\dagger}$Interdisciplinary Centre for Security, Reliability and Trust (SnT), University of Luxembourg, Luxembourg} \\
\textit{${}^{\ddagger}$SES S.A., Luxembourg}{\vspace{-3mm}}}}

\maketitle
%\IEEEpeerreviewmaketitle

\begin{abstract}
This paper presents a centralized framework for optimizing the joint design of beam placement, power, and bandwidth allocation in an MEO satellite constellation to fulfill the heterogeneous traffic demands of a large number of global users. The problem is formulated as a mixed integer programming problem, which is computationally complex in large-scale systems. To overcome this challenge, a three-stage solution approach is proposed, including user clustering, cluster-based bandwidth and power estimation, and MEO-cluster matching. A greedy algorithm is also included as a benchmark for comparison. The results demonstrate the superiority of the proposed algorithm over the benchmark in terms of satisfying user demands and reducing power consumption.
\end{abstract}
%\vspace{-1mm}

% \fi

\vspace{-3mm}

\section{Introduction}
%\cite{MIT_work},

Satellite communications have become an integral part of modern communication networks, providing seamless global connectivity for a wide range of applications such as broadband internet, remote sensing, and global positioning. 
With the increasing demand for high data rates and low latency communications, Medium Earth Orbit (MEO) constellations have emerged as a promising solution to enhance the capacity and coverage of satellite communication systems \cite{Al-Hraishawi_CST2022}. 
In addition, these Non-geostationary (NGSO) constellations are equipped with next-generation payload and antenna technologies including beam placement and  radio resource management (RRM) which can adapt their multi-beam capabilities to the global services \cite{Su2019}.
However, the design of such systems is a challenging task due to the large scale of the constellation and the complexity of the system design \cite{Al-Hraishawi_CST2022}.
Motivated by the advances as well as the challenges of NGSO satellite constellation, the goal of this work is to minimize the power consumption in MEO constellations while meeting the Quality of Service (QoS) requirements of all users.

The beam placement and RRM in satellite communication systems are two critical aspects that have received extensive attention in the literature. A variety of objectives have been proposed for beam placement, including minimizing half power beam footprint \cite{Pachler_IJSCN2021}, managing inter-beam interference \cite{Liu_GC2020}, and balancing inter-beam traffic loads fairly \cite{Phuc_MeditCon2022}. The diverse and changing quality of service requirements for different payloads \cite{Xia_Access2019,Hung_WSA23} have motivated the development of dynamic RRM solutions for multi-beam satellite communication systems that can accommodate irregular demands from multiple users \cite{VuHa_TWC22,VuHaGC2022}. However, these solutions can be challenging to implement in large-scale systems consisting of multiple MEO payloads serving thousands of ground users.

In response to recent advancements in MEO satellite constellations, this work focuses on optimizing beam placement, user-beam mapping, and RF resource allocation for global coverage. To address this challenge, a centralized power-minimization framework is proposed for joint beam placement, power, and bandwidth allocation design in MEO constellations. The framework consists of three stages: user clustering, cluster-related bandwidth and power estimation, and MEO-cluster matching. In the first stage, a novel clustering algorithm is proposed to divide users into a number of clusters, each of which can be served by one beam from an MEO. In the second stage, the optimal bandwidth and power consumed by each MEO to meet the demand of each of its corresponding clusters are estimated sequentially. Finally, in the third stage, the MEO-cluster assignment is optimized to minimize the total power consumption, based on the optimal bandwidth and power estimates from the previous stage. The proposed framework provides an efficient power-minimization approach for large-scale MEO constellation systems while ensuring that all users receive the desired Quality of Service (QoS). A greedy mechanism is also presented for comparison purposes. The numerical results show the efficiency and superiority of the proposed algorithm over other benchmark methods.

%*mention how we do beam design (ie placement) together with RRM. Which is difficult problem often done in sub problems. present our subproblems*

%*Furthermore we need to highlight the trade-off of assigning 1 beam per-user (many beams, many BW to be amplified, no multiplexing gain) and multiplexing users within a beam.*

%\vspace{-3mm}

%%%%%%%%%%%%%%%%%%%%%%%%%
\section{System Model}
%\begin{figure}[!t]
%	\centering
%	\includegraphics[width=70mm]{Figs/MEO_system.eps}
%	\caption{A multi-beam MEO-constellation satellite communication system.}
%	\label{MEO_constel}
 %   \vspace{-5mm}
%\end{figure}
Let us consider a SATCOM constellation  system consisting of $N$ MEO satellites serving $K$ users located on the earth's surface. These satellites move following the equator at the height of $h^{\sf{MEO}}$ while their longitudes, e.g.,  MEO $n$, over the time can be presented as $\theta_n(t)$, i.e., $\theta_n(t) \in [-180^\circ,180^\circ]$.
Let $\mathcal{N}$ and $\mathcal{K}$ be the sets of satellites and users, respectively.
Here, one assumes that users' demand traffic over a time window of $T$ time-slots (TSs) is available at SATCOM system, which is denoted as $D_k(t)$ for user $k$.
This constellation is controlled by a ground segment consisting of multiple gateways coordinated by a central controller including  an optimization module. 

\vspace{-2mm}

\subsection{Clustering-based Multi-Beam Satellite Communication}
In this system, each satellite can create multiple beams for serving users in its coverage area. However, activating a large number of beams is not always a beneficial choice. Moreover, serving multiple users within the same beam may be beneficial in terms of user multiplexing gain.
Therefore, one assumes that one beam may cover multiple users in an efficient manner instead of allocating one beam per user.
To develop a clustering framework, we introduce the binary variable $\{a_{m,k}\}$, which is defined as,
$a_{m,k} = 1$	if user $k \in \mathcal{C}_m$ 
and $a_{m,k} = 0$ otherwise, where $\mathcal{C}_m$ stands for the set of users belonging to cluster $m$. 
%Herein, the clusters can be scheduled according to the users' location and data demand. 
Additionally, the clustering solution is maintained unchanged during the time window for low-complexity design and one cluster can be served by only one MEO. 
Once cluster $m$ associates to MEO $n$, one specific beam focusing on this cluster will be generated by MEO $n$. 
Regarding cluster-MEO association, one introduces new variable $\{x_{m,k}(t)\}$ as,
\vspace{-1mm}
\beq \label{eq:meo_cl_match}
	x_{n,m} = 1 \text{ if $\mathcal{C}_m$ is served by MEO $n$ at TS $t$,}
 \vspace{-1mm}
\eeq
otherwise, $x_{n,m} = 0$. To ease the presentation, let us denote $\mathcal{M}$ as the set of all clusters, then, we have
\vspace{-2mm}
\beq
(C1): \quad \scaleobj{.8}{\sum_{\forall n}} x_{n,m}(t) \leq 1, \quad \forall m \in \mathcal{M}. 
\eeq

\vspace{-2mm}

\subsubsection{Beam Gain and Channel Gain Model}
\begin{figure}[!h]
	\centering
	\includegraphics[scale=0.23]{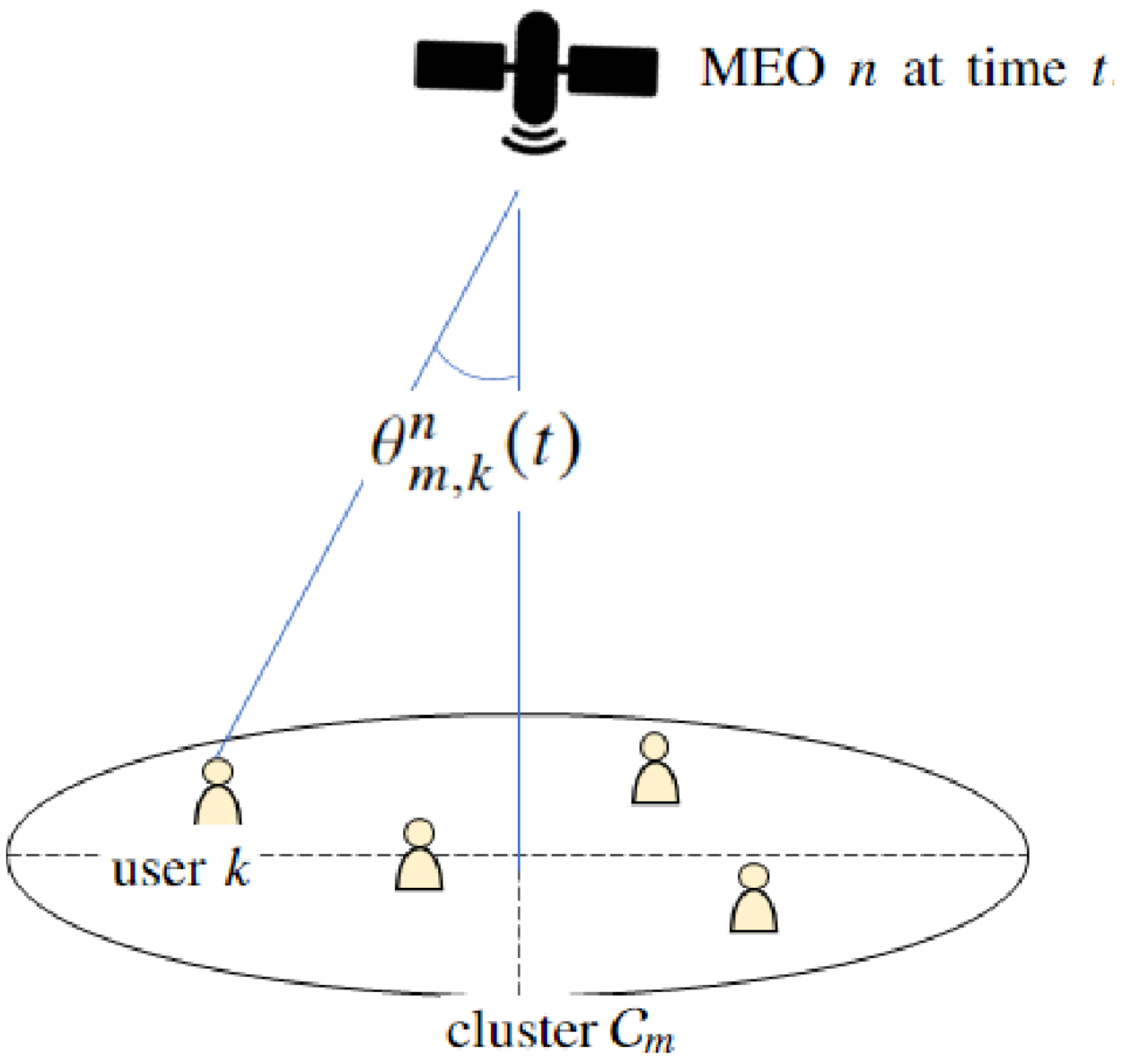}
	\caption{Geometry of the angle $\theta^n_{m,k}(t)$.}
	\label{theta_figure_fig}
    \vspace{-2mm}
\end{figure}

Let $\theta^n_{m,k}(t)$ be the angle between the beam center of MEO satellite $n$ at time $t$ to the user $k$ in cluster $\mathcal{C}_m$. The geometry of such angle is illustrated in Fig. \ref{theta_figure_fig}.
The beam radiation pattern function $g^n_{m,k}(t)$ between the beam $n$ and user $k$ can be formulated by following the 3GPP report \cite{3gpp2018study} as
	\vspace*{-0.1cm}
	\begin{equation} \label{eq:gbk}
		g^n_{m,k}(t) = g^{\max}\mathcal{G}(\theta^n_{m,k}(t)), \forall k \in \mathcal{C}_m, n \in \mathcal{N},
		\vspace*{-0.1cm}
	\end{equation}
	where $g^{\max}$ indicates the maximum gain and $\mathcal{G}(\theta^n_{m,k}(t))$ is the normalized beam pattern gain which is computed as
	\vspace*{-0.1cm}
	\begin{equation}\label{eq:gbktheta}
		\mathcal{G}(\theta) =  4\scaleobj{.8}{\left|{J_1\big(\scaleobj{.8}{\frac{2\pi}{\lambda}} r^{\sf{ant}} \sin(\theta)\big)}/({\scaleobj{.8}{\frac{2\pi}{\lambda}}r^{\sf{ant}}\sin (\theta))}\right|^2} , \mbox{if } 0 < \theta\leq \frac{\pi}{2}, 
		\vspace*{-0.1cm}
	\end{equation}
 and $\mathcal{G}(\theta) = 1$ if $\theta = 0$, where $J_1(\cdot)$, $\lambda$, $r^{\sf{ant}}$ represent the order-one Bessel function, the carrier wavelength, and the radius of the antenna’s circular aperture, respectively. Then, the channel gain from MEO $n$ to user $k$ can be described as
	$h^n_{m,k}(t) = g^n_{m,k}(t)P_{\sf{ls}}(d^n_k(t))$
where $P_{\sf{ls}}(d^n_k(t))$ represents the path-loss.
\vspace{-3mm}

\subsubsection{Achievable Rate}
Denote $P_k(t)$ and $B_k(t)$ as the transmission power and bandwidth allocated to user $k$ at time $t$. Then, the achievable rate of this user can be described as
\vspace{-1mm}
\beq
R_k(t) = \scaleobj{.8}{\sum_{\forall (n,m)}} a_{m,k} x_{n,m}(t) R^n_{m,k}(t),
\vspace{-1mm}
\eeq
where $R^n_{m,k}(t) = B_k(t)\log_2(1 + {P_k(t) h^n_{m,k}(t)}/{B_k(t) \sigma^2} )$ and $\sigma^2$ stands for the noise power per Hz at the receivers. Then, the constraint on users demand can be written as
\vspace{-1mm}
\beq
(C2): \quad  R_k(t) \geq D_k(t), \quad \forall (k,t).
\eeq

\vspace{-3mm}

\subsection{Power Consumption Model}
In our system, one assumes that the digital transparent processors (DTP) are equipped at all payloads based on which the high-power amplifiers (HPAs) and transponders can be switched off dynamically.
The total power consumption of all satellites in TS $t$ is mathematically formulated as
\beq
P^{\sf{tot}}(\mb{B}(t),\mb{P}(t)) = \scaleobj{.8}{\sum_{\forall n}} P_n^{\sf{hw}}(\mb{B}(t),\mb{P}(t)) + \scaleobj{.8}{\sum_{\forall k}} P_k(t),
\eeq
where $P_n^{\sf{hw}}(\mb{B}(t),\mb{P}(t))$ models the hardware power consumption
connected to all active beams of MEO $n$, which is consumed by various hardware components such as signal processing and HPA.
Herein, $P_n^{\sf{hw}}(\mb{B}(t),\mb{P}(t))$ is defined as
\beq
P^{\sf{hw}}_n(\mb{B}(t),\mb{P}(t)) = P^{\sf{DC}}_n(\mb{B}(t)) + P^{\sf{RF}}_n(\mb{P}(t))/\rho^{\sf{HPA}},
\eeq
where $\rho^{\sf{HPA}}$ is the DC to RF power efficiency of the antenna
element amplifiers. 
Additionally, $P^{\sf{RF}}_n(\mb{P}(t))$ denotes the RF transmission power of beam $n$ which is
computed as
\beq
P_n^{\sf{RF}}(\mb{P}(t)) = \scaleobj{.8}{\sum_{\forall (m,k)}} a_{m,k} x_{n,m}(t) P_k(t),
\eeq
%where $B^{\sf{RTN}}$ and $P^{\sf{RF}}_{\sf{max}}$ represent as the bandwidth of the return link
%and the maximal RF power on the forward link antenna, $B_n(t)$ and $B^{\sf{tot}}_n$ denote the utilized and maximum BW of MEO $n$, respectively.
The DC consumed power according to the onboard signal processing, denoted by $P^{\sf{DC}}_n(\mb{B}(t))$, is given as
\beq
P^{\sf{DC}}_n(\mb{B}(t)) = {P^{\sf{DC}}_{\sf{tot}} B_n(t)}/{B^{\sf{tot}}},
\eeq
where $P^{\sf{DC}}_{\sf{tot}}$ represents the max DC power and $B^{\sf{tot}}_n$ í the max BW of MEO.
Herein, $B_n(t)$ is estimated as
\beq
B_n(t) = \scaleobj{.8}{\sum_{\forall (m,k)}} a_{m,k} x_{n,m}(t) B_k(t).
\eeq
Then, if $B^{\sf{tot}}_n$ is the same for all $n$, which is denoted as $B^{\sf{tot}}$, the total power consumption of all MEOs can be estimated as
\beq
P^{\sf{tot}}(\mb{B}(t),\mb{P}(t)) = \scaleobj{.8}{\sum_{\forall k}} \left( \scaleobj{.8}{\dfrac{\rho^{\sf{HPA}}+1}{\rho^{\sf{HPA}}}}P_k(t) + \scaleobj{.8}{\dfrac{P^{\sf{DC}}_{\sf{tot}}}{B^{\sf{tot}}}} B_k(t) \right).
\eeq

\vspace{-5mm}

\subsection{Problem Formulation}
This work aims to develop a centralized power-minimization framework of joint beam placement, power and BW allocation for MEO-constellation which 
is stated as
	\begin{subequations} \label{MEO-Clus_Prob}
		\begin{eqnarray} 
			\hspace{-0.8cm}&\underset{\mb{a},\mb{x},\mb{B},\mb{P}}{\min}& \hspace{-0.2cm} {\scaleobj{.8}{\sum_{\forall t}}} P^{\sf{tot}}(\mb{B}(t),\mb{P}(t)) \label{obj_func_MinPower}\\
			\hspace{-0.8cm}&\text{s.t. }&  \text { constraints $(C1), (C2)$,} \nonumber \\
			\hspace{-0.8cm}& &	\hspace{-0.2cm} (C3): \; \scaleobj{.8}{\sum_{\forall m}} x_{n,m}(t) \scaleobj{.8}{\sum_{k \in \mathcal{C}_m}} a_{m,k} B_k(t) \leq B^{\sf{tot}}, \; \forall n, \label{cnt2} \\
			\hspace{-0.8cm}&& \hspace{-0.2cm} (C4): \; \scaleobj{.8}{\sum_{\forall m}} x_{n,m}(t) \scaleobj{.8}{\sum_{k \in \mathcal{C}_m}} a_{m,k} P_k(t) \leq P^{\sf{RF}}_{\sf{max}},  \; \forall n, \label{cnt3}
		\end{eqnarray}
	\end{subequations}
where $(C3)$ and $(C4)$ stand for the constraints on the limited power transmission and BW at every payload.
\begin{remark}
This problem is a mixed integer programming which is well-known as NP-hard. Additionally, solving it optimally becomes more challenging in a large-scale global setting with the very huge number of users. 
\end{remark}

\section{Solution Approach}
In this section, a three-stage solution approach is proposed to cope with the challenging problem given in \eqref{MEO-Clus_Prob}.
In particular, this solution consists of three stages:
\begin{enumerate}
    \item User clustering: Scheduling users into separated groups each of which can be served by one beam from an MEO.
    \item Cluster-related BW and Power Estimation: Estimating the optimal BW and power by an assigned-to-serve MEO to meet users' demand in a cluster.
    \item MEO-Cluster Matching: Optimizing the MEO-serving-cluster assignment to minimize the total power consumption according to the optimal BW and power estimated in the second stage.
\end{enumerate}

\vspace{-2mm}

\subsection{User Clustering}
\subsubsection{Potential User Clustering Matrix}
Assuming the coverage of every generated beam can be defined as the footprint of $3$-dB loss from the beam center. The users should be grouped into clusters so that the angle from a satellite to two arbitrary users in a cluster must be not greater than the beam width.
Let $\mb{U} \in \mathbb{R}^{K \times K}$ be the adjacency matrix whose $(k,\ell)$-th element, denoted by $[\mb{U}]_{k,\ell}$, is defined as follows
\beq \label{eq:U_kl}
	[\mb{U}]_{k,\ell} = 
	\left\lbrace \begin{array}{*{10}{l}}
1, & \text{if } \theta^{\sf{max}}_{k,\ell} < \theta^{\sf{beam}},  \\
0, & \text{otherwise},
\end{array}
\right. 
\eeq
where $\theta^{\sf{max}}_{k,\ell}$ represents the maximum angle from a point on satellite orbit to users $k$ and $\ell$, and $\theta^{\sf{beam}}$ is the beam width. 
Herein, $\theta^{\sf{max}}_{k,\ell} < \theta^{\sf{beam}}$ implies that users $k$ and $\ell$ can be served by one beam; hence, they can be grouped into one cluster.
Let $\mathcal{O}^{\sf{MEO}}$ be the MEO orbit line, then $\theta^{\sf{max}}_{k,\ell}$ can be defined as
\beq
\theta^{\sf{max}}_{k,\ell} = \underset{X \in \mathcal{O}^{MEO}} \max \measuredangle{ kX\ell},
\eeq
where $\measuredangle{ kX\ell}$ denotes the angle separating users $k$ and $\ell$ from $X$'s point of view and $\mathcal{O}^{MEO}$ represents the orbit of all MEOs.
\begin{remark}
It is worth noting that the angle from any MEO to two users $k$ and $\ell$ is always less than $\theta^{\sf{max}}_{k,\ell}$. 
\end{remark}

\subsubsection{User's Required MEO Bandwidth Estimation}
This section aims to estimate amount of MEO BW required to serve a specific ground user.
Consider user $k$ with demand $D_k(t)$.
%The BW allocated to this user must be large enough to satisfy its peak demand, which denote as $D^{\sf{max}}_k$, i.e., $D^{\sf{max}}_k = \max_t D_k(t)$.
Note that user $k$ may be located in a cluster that be covered and served by an activated beam from MEO $n$. The coverage of that beam is defined by the footprint of $3$-dB loss from the beam center.
Here, we consider the worst case that user $k$ laying on the boundary of that beam than, $g^{\sf{bound}} = g^{\sf{max}}/2$.
And the distance in the worst case can be defined as
\beq
\bar{l}_k = \max_t \min_n l^n_{k}(t),
\eeq
where $l^n_{k}(t)$ stands for the distance between user $k$ and MEO $n$ at time $t$.
Then, the estimated BW which should be allocated for user $k$ over the time can be presented as, $\bar{B}_k(t)$, where
\beq
\bar{B}_k(t) \log_2\left(1+\scaleobj{.8}{\dfrac{P^{\sf{max}}_{\sf{beam}} g^{\sf{bound}} P_{\sf{ls}}(\bar{l}_k)}{\bar{B}_k(t) \sigma^2}}\right) = D_k(t).
\eeq
Based on $\bar{B}_k(t)$, the simple clustering algorithm is presented in what follows.

\subsubsection{Time-Window-based Matching Efficiency Factor}
In this work, one assumes that the cluster results cannot be changed during a time window of $[0,T]$.
Hence, users should be grouped into clusters so that the total required bandwidth of each cluster does not vary too much and also its average is not very far from the peak value during the time window.
To do so, we present a new time-window-based matching efficiency factor as follows.
Let $\mathcal{S}$ be a set of some arbitrary users. 
The time-window-based matching efficiency factor corresponding to this set can be described as 
\beq
E(\mathcal{S}) = \scaleobj{.8}{ {\int_0^T\sum_{k \in \mathcal{S}} }\bar{B}_k(t)dt}/{(T\times \bar{B}^{\sf{max}}(\mathcal{S}))} ,
\eeq
where $\bar{B}^{\sf{max}}(\mathcal{S}) = \underset{t\in[0,T]}{\max}\sum_{k \in \mathcal{S}}\bar{B}_k(t)$.
%Denote $B^{\sf{beam}}_{\sf{max}}$ as the maximum bandwidth of one beam and $\mathcal{K}$ as the set of all users.

\subsubsection{A Simple Clustering Algorithm}
In what follows, a greedy clustering algorithm is proposed to separate all users into different clusters in a manner that one beam can be placed to serve all users in a cluster and the users are grouped so that the matching efficiency factor of the corresponding cluster is as high as possible. In particular, the proposed clustering approach is summarized in Algorithm~\ref{P2_alg:2}.

\begin{algorithm}[!t]
\footnotesize
	\caption{\textsc{Proposed Clustering Algorithm}}
	\label{P2_alg:2}
	%\algsetup{indent=1.5em}
	\begin{algorithmic}[1]
		\STATE \textbf{Initialize:} Set $\mb{U}^{\sf{temp}} = \mb{U}$ and cluster index as $id = 0$.
  %\begin{enumerate}[label = {1-\alph*:}]
			%\item Set $\mb{U}^{\sf{temp}} = \mb{U}$.
			%\item Set cluster index as $id = 0$.
		%\end{enumerate}  
		\FOR{$k=1$ to $K$}
		\IF{$[\mb{U}^{\sf{temp}}]_{k,k} = 1$}
	    \STATE Update $id = id +1$ and set $id$-th cluster as $\mathcal{C}_{id} = \{k\}$.
	    \STATE Update $\mb{U}^{\sf{temp}}$ by setting all elements on its $k$-th column to zeros. 
	    \STATE Define the set of potential users which can be added into this cluster as $\mathcal{U}_{id} = \lbrace \ell | [\mb{U}^{\sf{temp}}]_{k,\ell} = 1\rbrace$.
	    \WHILE{$\bar{B}^{\sf{max}}(\mathcal{C}_{id})< B^{\sf{beam}}_{\sf{max}}$}
	    \STATE Define the next user to add to $id$-th cluster as 
	    \beq \label{eq:next_user}
	    \hspace{-8mm} \ell^{\prime}=\arg\underset{\ell \in \mathcal{U}_{id}}{\max} E(\mathcal{C}_{id}\cup \{\ell\}) \text{ s.t. } \bar{B}^{\sf{max}}(\mathcal{C}_{id}\cup \{\ell\}) \leq B^{\sf{beam}}_{\sf{max}}.
	    \eeq
	    \vspace{-2mm}
	    \IF{One exists a such user $\ell^{\prime}$ as in \eqref{eq:next_user}}
	    \STATE Add user $\ell^{\prime}$ into $\mathcal{C}_{id}$ as $\mathcal{C}_{id}=\mathcal{C}_{id}\cup \{\ell\}$.
	    \STATE Set all elements on $\ell^{\prime}$-th column of $\mb{U}^{\sf{temp}}$ to zeros.
	    \STATE Update $\mathcal{U}_{id} = \mathcal{U}_{id} \cap \lbrace m | [\mb{U}^{\sf{temp}}]_{\ell^{\prime},m} = 1\rbrace$.
	    \ELSE
	    \STATE Break (Stop WHILE loop).
	    \ENDIF
	    \ENDWHILE
	    \ENDIF
		\ENDFOR
	\end{algorithmic}
 \normalsize  
 \end{algorithm}

 \vspace{-2mm}

\subsection{MEO-Cluster BW and Power Estimation}
 \vspace{-1mm}
\subsubsection{Beam Center Allocation}
The beam center of each cluster can be defined simply based on the coordinates and demands of all users in cluster $\mathcal{C}_m$ as follows
\beq
\mb{o}(\mathcal{C}_m) \! = \! \scaleobj{.8}{\frac{\sum_{k \in \mathcal{C}_m} \! ( \eta + D_k^{\sf{max}}) \mb{o}_k}{\sum_{k \in \mathcal{C}_m} \! (\eta + D_k^{\sf{max}})} } \! = \! \scaleobj{.8}{\frac{\sum_{\forall k} a_{n,k} ( \eta + D_k^{\sf{max}}) \mb{o}_k}{\sum_{\forall k} a_{n,k} (\eta + D_k^{\sf{max}})}},
\eeq
where $\eta$ is the calibrating factor, $D_k^{\sf{max}} = \max_t D_k(t)$, and $\mb{o}_k$ be the coordinate of user $k$. 
%It is worth noting that the beam center location is also an interesting research topic that is not covered in this work.
\subsubsection{Required BW and Power Estimation}
Assume that MEO $n$ having coordinate $\mb{o}^{\sf{MEO}}_n(t)$ is assigned to serve cluster $\mathcal{C}_m$ in TS $t$.
According to $\mb{o}(\mathcal{C}_m)$ and $\mb{o}^{\sf{MEO}}_n$, $\theta^n_{m,k}(t)$, so-called the angle from beam center to user $k$ in TS $t$, can be defined.
Based on that, the corresponding channel gain $h^n_{m,k}(t)$ can be estimated.
Let $B_{n,k}(t)$ and $P_{n,k}(t)$ be the assigned BW and transmission power to serve user $k$ by MEO $n$. Then, $B_{n,k}(t)$ and $P_{n,k}(t)$ can be estimated by solving the following problem.
	\begin{subequations} \label{MINPOW_CLuster}
 \vspace{-2mm}
		\begin{eqnarray} 
			\hspace{-0.8cm}&\underset{\{B_{n,k}(t), P_{n,k}(t)\}}{\min}& \hspace{-0.4cm} \Phi_{n,m}(t) \! = \! \scaleobj{.8}{\sum_{\forall k}} \!\! \left( \scaleobj{.7}{\dfrac{\rho^{\sf{HPA}}+1}{\rho^{\sf{HPA}}}}P_{n,k}(t) + \scaleobj{.7}{\dfrac{P^{\sf{DC}}_{\sf{tot}}}{B^{\sf{tot}}}} B_{n,k}(t) \right) \label{obj_func_MINPOW_Clu}\\
			\hspace{-0.8cm}&\text{s.t. }&  \hspace{-1cm} B_{n,k}(t) \! \log_2 \! \left(1 \! + \! \scaleobj{.7}{\dfrac{P_{n,k}(t) h^n_{m,k}(t)}{B_{n,k}(t) \sigma_k}}\right) \! \geq \! D_k(t), \forall k \! \in \! \mathcal{C}_m. \label{cnt1}
		\end{eqnarray}
	\end{subequations}
\begin{theorem}
Problem \eqref{MINPOW_CLuster} is convex
\end{theorem}
\begin{IEEEproof}
As can be observed, $B_{n,k}(t) \log_2\left(1+\scaleobj{.7}{\dfrac{P_{n,k}(t) h^n_{m,k}(t)}{B_{n,k}(t) \sigma_k}} \right)$ is a joint concave function respected to both $B_{n,k}(t)$ and $P_{n,k}(t)$. Additionally, the objective function is in the linear form of the variables. Thus, problem \eqref{MINPOW_CLuster} must be convex. 
\end{IEEEproof}
Hence, it can be solved optimally efficiently by employing several convex optimization tools such as CVX, Gurobi \cite{Hung_ICCWks23}.
\begin{remark}
Note that only the clusters located inside each MEO's field of view (FoV) are considered for estimating the required BW and power. Herein, the FoV of every MEO is defined by its attitude and minimum elevation angle \cite{Angeletti_Access20,VuHa_VTCFall22}.
\end{remark}

\vspace{-3mm}

\subsection{MEO-Cluster Matching Design}
Denote $B_{n,k}^{\star}(t)$ and $P_{n,k}^{\star}(t)$ as the solution of problem~\eqref{MINPOW_CLuster} for MEO $n$ and user $k$.
Then, the MEO-Cluster matching problem can be stated as
	\begin{subequations} \label{MEO-Clus_Prob}
		\begin{eqnarray} 
			\hspace{-0.8cm}&\underset{\{x_{n,m}(t)\}'s}{\min}& \hspace{-0.2cm} {\scaleobj{.8}{\sum \limits_{\forall (n,m)}}} x_{n,m}(t) \Phi_{n,m}^{\star}(t) \label{obj_func_MEOClus}\\
			\hspace{-0.8cm}&\text{s.t. }&  \hspace{-0.2cm} \scaleobj{.8}{\sum_{\forall m}} x_{n,m}(t) \scaleobj{.8}{\sum_{k \in \mathcal{C}_m}} B_{n,k}^{\star}(t) \leq B^{\sf{max}}_n, \; \forall n, \label{cnt2} \\
			\hspace{-0.8cm}&& \hspace{-0.2cm} \scaleobj{.8}{\sum_{\forall m}} x_{n,m}(t) \scaleobj{.8}{\sum_{k \in \mathcal{C}_m}} P_{n,k}^{\star}(t) \leq P^{\sf{max}}_n,  \; \forall n, \label{cnt3}
		\end{eqnarray}
	\end{subequations}
where $\Phi_{n,m}^{\star}(t) =  \sum_{\forall k} \scaleobj{.8}{\left( \dfrac{\rho^{\sf{HPA}}+1}{\rho^{\sf{HPA}}}P_{n,k}^{\star}(t) + \dfrac{P^{\sf{DC}}_{\sf{tot}}}{B^{\sf{tot}}} B_{n,k}^{\star}(t) \right)}$. This is an integer linear programming in general which can be solved efficiently by using an \textit{``off-the-shelf''} optimization toolbox, or by employing a relaxation and projection approach introduced in \cite{VuHa_Access17,Sanjabi_TSP14,VuHa_TWC18}. 
%,VuHa_Access17}.
Then, the proposed solution approach is summarized in 
Algorithm~\ref{P2_alg:1}.
\begin{remark}
Note that the cross interference among all clusters is omitted in this work thanks to the narrow beam-width beam patterns of the satellite and that MEO can allocate different frequency bands to adjacent beams \cite{Angeletti_Access20,VuHa_GCWks22}.
\end{remark}

\begin{algorithm}[!t]%\leesize
\caption{\textsc{Proposed Algorithm}}
\label{P2_alg:1}
%\algsetup{indent=1.5em}
%\scriptsize
\footnotesize
\begin{algorithmic}[1]
	%\REQUIRE Maximum iteration number $N$, tolerance $\epsilon$.
	\STATE Employing Algorithm~\ref{P2_alg:2} to group all users into clusters.
    \STATE Defining $\mathcal{M}$, $\mathcal{C}_m$'s,and $\mb{o}_m$'s.
	\FOR{TS $t = 1:T$} 
    \FOR{MEO $n \in \mathcal{N}$}
    \FOR{Cluster $m \in \mathcal{M}$}
    \IF{$\mb{o}_m$ locates inside the coverage of MEO $n$}
    \STATE Solving problem \eqref{MINPOW_CLuster} for MEO $n$ and $\mathcal{C}_m$ to obtain the corresponding $B_{n,k}(t)$'s and $B_{n,k}(t)$'s.
    \ELSE
    \STATE Setting the corresponding $B_{n,k}(t)$'s and $B_{n,k}(t)$'s to \textit{+Inf}.
    \ENDIF
    \ENDFOR
    \ENDFOR
    \STATE Solving problem \eqref{MEO-Clus_Prob} based on $B_{n,k}^{\star}(t)$'s and $P_{n,k}^{\star}(t)$'s to get the matching results.
    \ENDFOR
\end{algorithmic}
\normalsize
\end{algorithm}

\begin{algorithm}[!t]%\leesize
\caption{\textsc{Low Complexity Design}}
\label{P2_alg:4}
%\algsetup{indent=1.5em}
\footnotesize
\begin{algorithmic}[1]
	\STATE Determine maximum cluster distance  $d^{\sf{clu}}= 2h^{\sf{MEO}}\tan{(\theta^{\sf{beam}})}$. 
	\STATE Exploiting the method in \cite{MaNing_JIoT2022} to cluster all users.
    \FOR{TS $t = 1:T$} 
     \STATE Every cluster is connected to its nearest MEO satellite.
    \STATE Solving problem \eqref{MINPOW_CLuster} for each MEO to obtain the BW and power.
    \ENDFOR
\end{algorithmic}
\normalsize
\end{algorithm}

\vspace{-3mm}

\subsection{Greedy Solution Approach}
As a point of comparison, we also created a greedy solution approach. This involves using the Voronoi clustering algorithm from \cite{MaNing_JIoT2022} to divide all users into clusters. The maximum distance for each cluster is calculated based on the beam width, $\theta^{\sf{beam}}$, and orbit height, $d^{\sf{clu}} = 2h^{\sf{MEO}}\tan{(\theta^{\sf{beam}})} $. After the clustering is done, each cluster is assigned to its closest MEO. Finally, the bandwidth and power transmission of all users are optimized by solving problem \eqref{MINPOW_CLuster} for each MEO. The whole process is outlined in Algorithm~\ref{P2_alg:4}.

\section{Numerical Results}
\begin{figure}[!t]
	\centering
	\includegraphics[width=80mm]{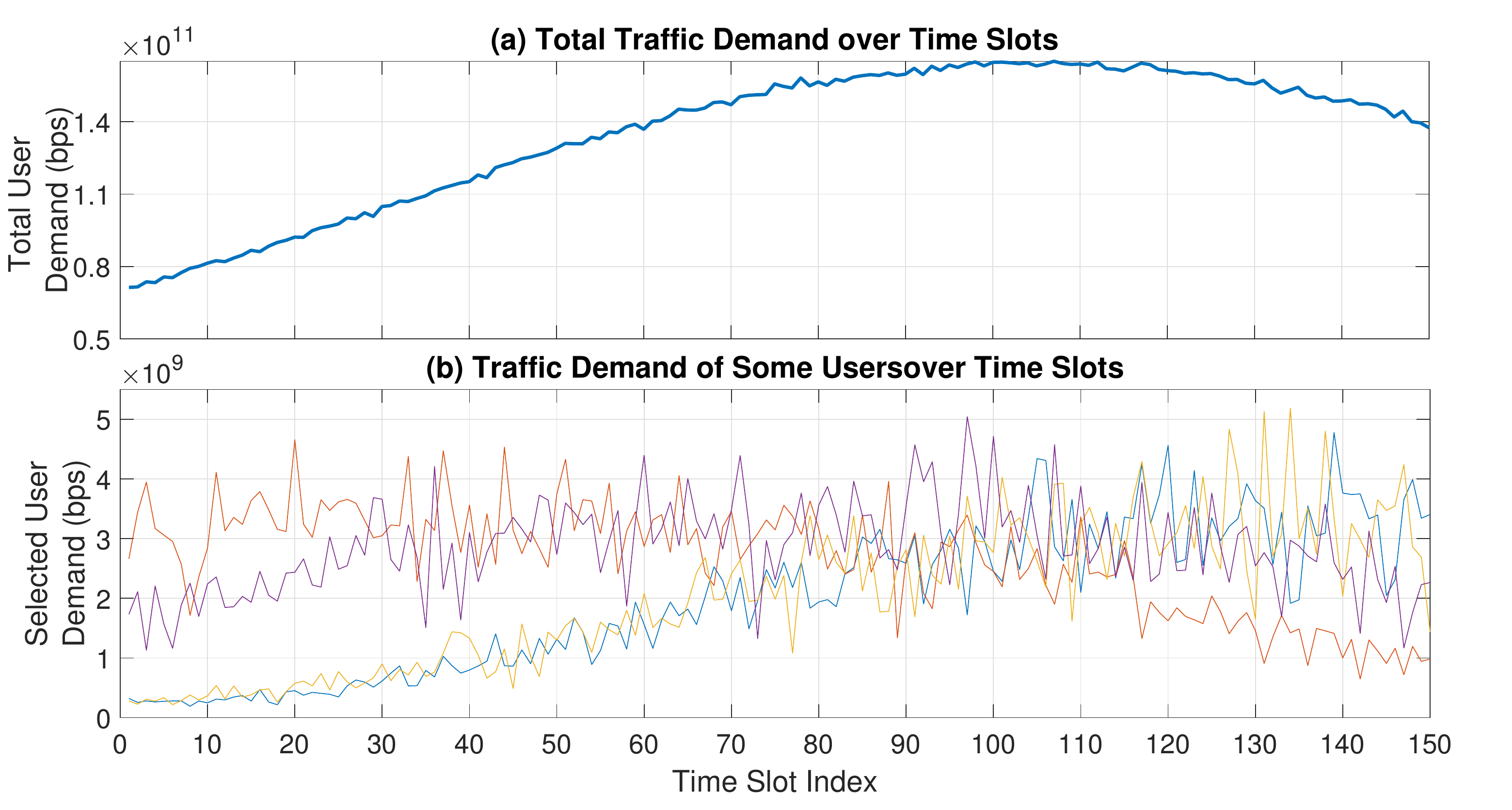}
 \vspace{-1mm}
	\caption{A realization of user demand over $150$ TSs.}
	\label{User_150TS} 
\vspace{-3mm}
\end{figure}

In this section, some simulations conducted for an MEO constellation of 15 payloads orbiting at an altitude of $8062$ km will be provided. The simulations were run over a time frame of $150$ TSs, each lasting 6 minutes. In this simulation, the users were placed in proximity to major cities each of which contains around $40$ users. Then, there was a total of $1112$ users over the world. The user demands were randomly generated based on sinusoidal patterns with peak and off-peak periods. The total demand of all users over the $150$ TSs is displayed in Fig.~\ref{User_150TS}. The parameters for the simulation are listed in Table~\ref{setting_paras} and preliminary numerical results are also provided. 

\begin{table}[!t]
\scriptsize
%\footnotesize
	\centering
%	\captionsetup{font=footnotesize}
	\caption{\textsc{Simulation Parameters (Provided by ESA)}}
 \vspace{-2mm}
	\begin{tabular}{l| r}
		\hline
		Parameter & Value \\
		\hline\hline
	%	Number of Monte Carlo simulations												& $50$		\\
		Forward link carrier frequency										& $19-21.5$~GHz \\
		MEO attitude 	& $8062$~km\\
		Earth radius & $6378$~km \\
		$P^{\sf{RF}}_{\sf{max}}$ & $800$~W \\
		$\rho^{\sf{HPA}}$ & $0.6$ \\
		$P^{\sf{DC}}_{\sf{tot}}$ & $5000$~W \\
		$B^{\sf{tot}}$ 	& $2500$~MHz \\ 
		Minimum satellite elevation angle					& $5$~degrees\\	
		Number of simulated users around a big city					& $20-50$ \\
		User terminal antenna gain												& $41.45$~dBi\\
		Temperature at user terminals													& $224.5$~K\\
		Channel Model							& Refer to~\cite{Angeletti_Access20}   \\
	\hline
	\end{tabular} \label{setting_paras} \vspace{-3mm}
	\end{table}

\begin{figure}[!t]
	\centering
	\includegraphics[width=70mm]{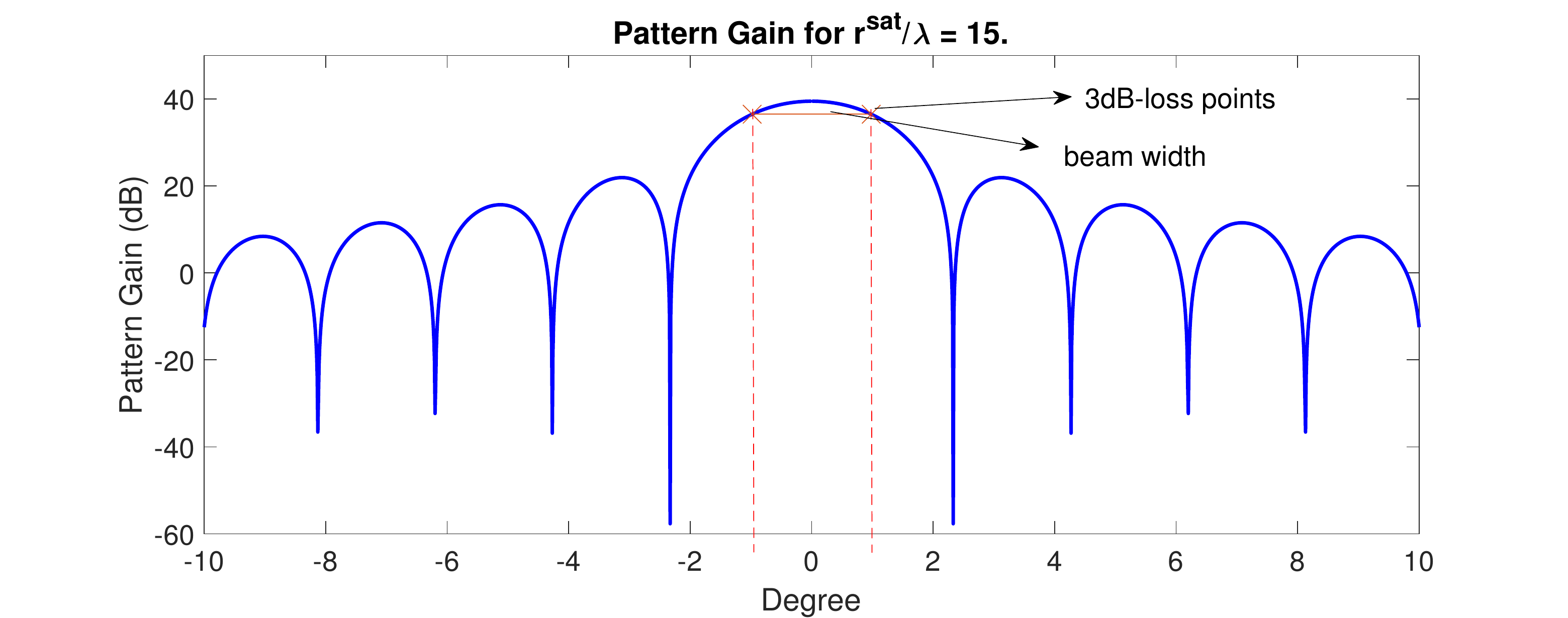}
	\caption{An example of beam pattern gain for $r^{\sf{ant}}/\lambda=15$.}
	\label{Beam_Width_Example}
 \vspace{-3mm}
\end{figure}

In Table~\ref{beamwidth_radius}, we determine the beam width based on the radius of the antennas installed on the payload. By using the normalized beam pattern gain given in \eqref{eq:gbktheta}, the beam width is calculated as the $3$dB loss angle, which is defined as $\theta^{\sf{beam}} = 2 \theta \big\vert_{\mathcal{G}(\theta) = 0.5}$. Figure~\ref{Beam_Width_Example} shows an example of the beam pattern gain for $r^{\sf{ant}}/\lambda=15$. It's worth mentioning that the beam width of $1.96^{\circ}$ for $r^{\sf{ant}}/\lambda=15$ is also utilized for the antenna configuration in all subsequent simulations.

\begin{table}[!t]
%\scriptsize
\scriptsize
	\centering
%	\captionsetup{font=footnotesize}
	\caption{\textsc{Beam Width and Radius of Satellite Antenna}}
 \vspace{-2mm}
	\begin{tabular}{| c | c | c | c | c |}
		\hline
        $r^{\sf{ant}}/\lambda$ & $5$ & $10$ & $15$ & $20$ \\
        \hline
        $r^{\sf{ant}}$ & $7.89$ (cm) & $15.79$ (cm) & $23.68$ (cm) & $31.58$ (cm) \\
        \hline
        $\theta^{\sf{beam}}$ & $5.88^{\circ}$ & $2.94^{\circ}$ & $1.96^{\circ}$ & $1.46^{\circ}$ \\
		\hline
  $d^{\sf{clu}}$ & $828.09$ (km) & $413.77$ (km) & $275.81$ (km) & $205.44$ (km)\\
		\hline
	\end{tabular} \label{beamwidth_radius}  \vspace{-3mm}
	\end{table} 

 \begin{table}[!t]
%\scriptsize
\scriptsize
	\centering
%	\captionsetup{font=footnotesize}
	\caption{\textsc{Cluster Results}}
 \vspace{-2mm}
	\begin{tabular}{| c | c | c | c |}
		\hline
        Method & Number of Cl. & Min/Max Cl. Size & Avg. Cl. Size \\
        \hline
        Proposed Alg. & $386$ & $1$/$10$ & $2.88$ \\
        \hline
        Voronoi \cite{MaNing_JIoT2022} & $477$ & $1$/$8$ & $2.33$ \\
        \hline
	\end{tabular} \label{clustering_result}  \vspace{-3mm}
\end{table} 

\begin{figure}[!t]
	\centering
	\includegraphics[width=90mm]{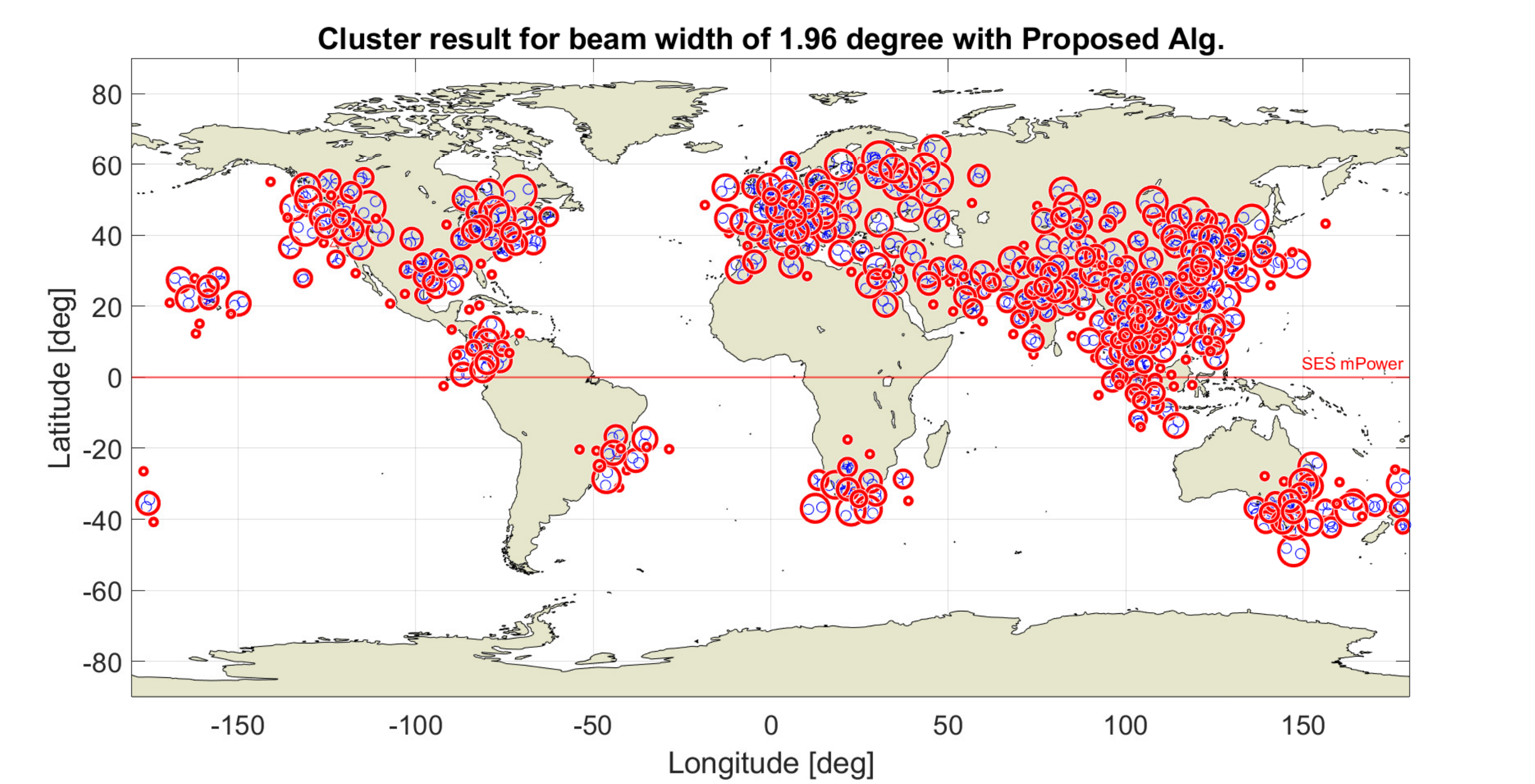}
	\caption{Proposed Alg. based Clustering result with angle threshold $1.96^{\circ}$.}
	\label{Cluster_Example196}  \vspace{-3mm}
\end{figure}

\begin{figure}[!t]
	\centering
	\includegraphics[width=90mm]{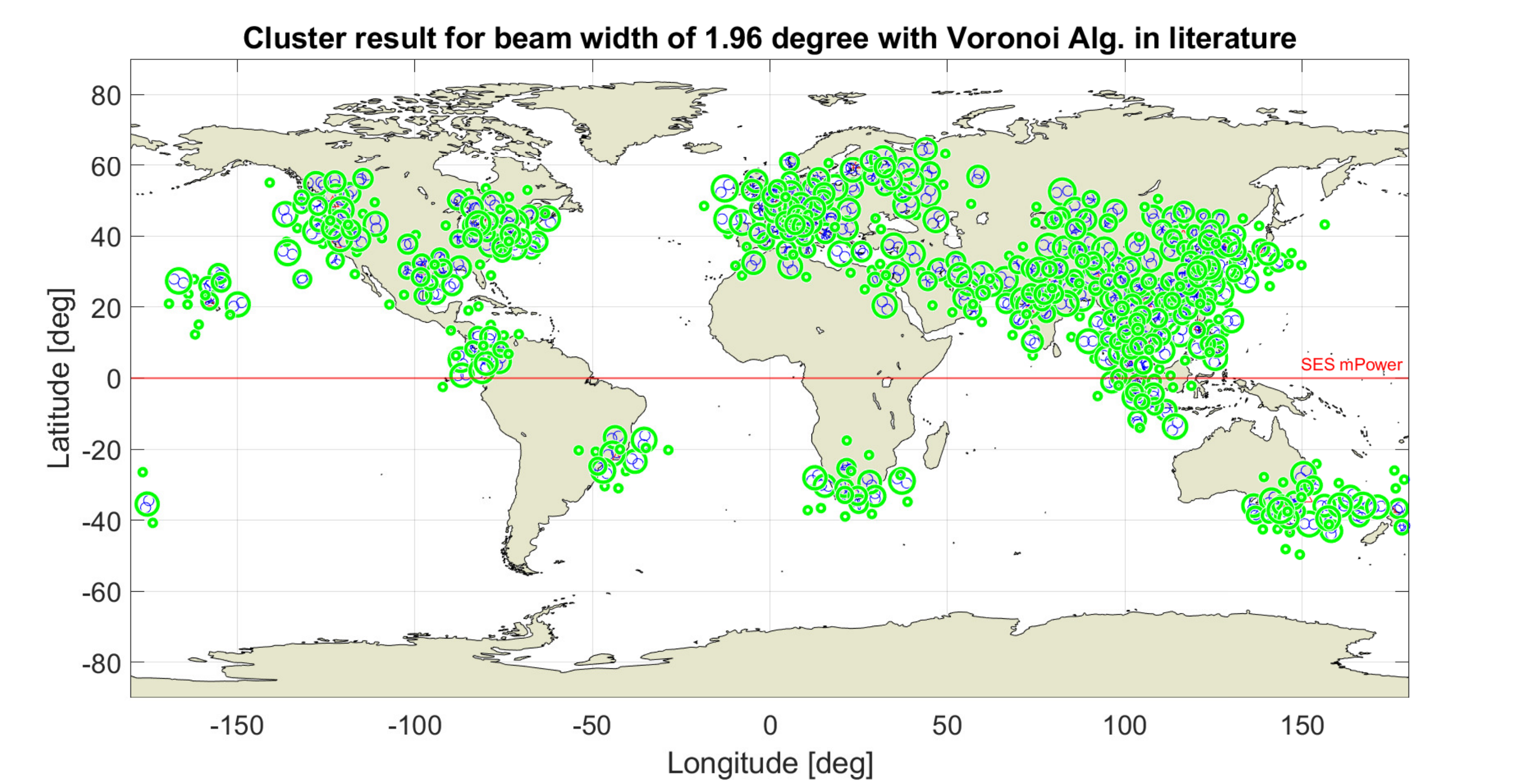}
	\caption{Voronoi Alg. \cite{MaNing_JIoT2022} based Clustering result with angle threshold $1.96^{\circ}$.}
	\label{Cluster_Example588}  \vspace{-3mm}
\end{figure}

Figs.~\ref{Cluster_Example196} and \ref{Cluster_Example588} demonstrate the clustering results produced by using our proposed method in Algorithm~\ref{P2_alg:2} and the distance-based Voronoi method from \cite{MaNing_JIoT2022}, respectively. These figures show the clustering results based on satellite positions and user demands in TS $1$. The information is further outlined in Table~\ref{clustering_result}. Our proposed approach can be seen to generate a fewer number of beams compared to the other method.

\begin{table}[!t]
\scriptsize
%\footnotesize
	\centering
%	\captionsetup{font=footnotesize}
	\caption{\textsc{Cluser-MEO Matching Results}}
 \vspace{-2mm}
	\begin{tabular}{| c | c | c | c | }
	\hline
       Method & Min/Max Clus./MEO& Ser. Clus. Ratio & Satis. User No.\\
        \hline
       Proposed Alg. & $14$/$55$ & $380/386$ & $1089/1112$  \\
       \hline
        Greedy Alg. & $7$/$77$ & $397/477$ & $889/1112$\\
       \hline
	\end{tabular} \label{cluster_MEO_Mathcing_result} \vspace{-3mm}
\end{table} 

\begin{figure}[!t]
	\centering
	\includegraphics[width=90mm]{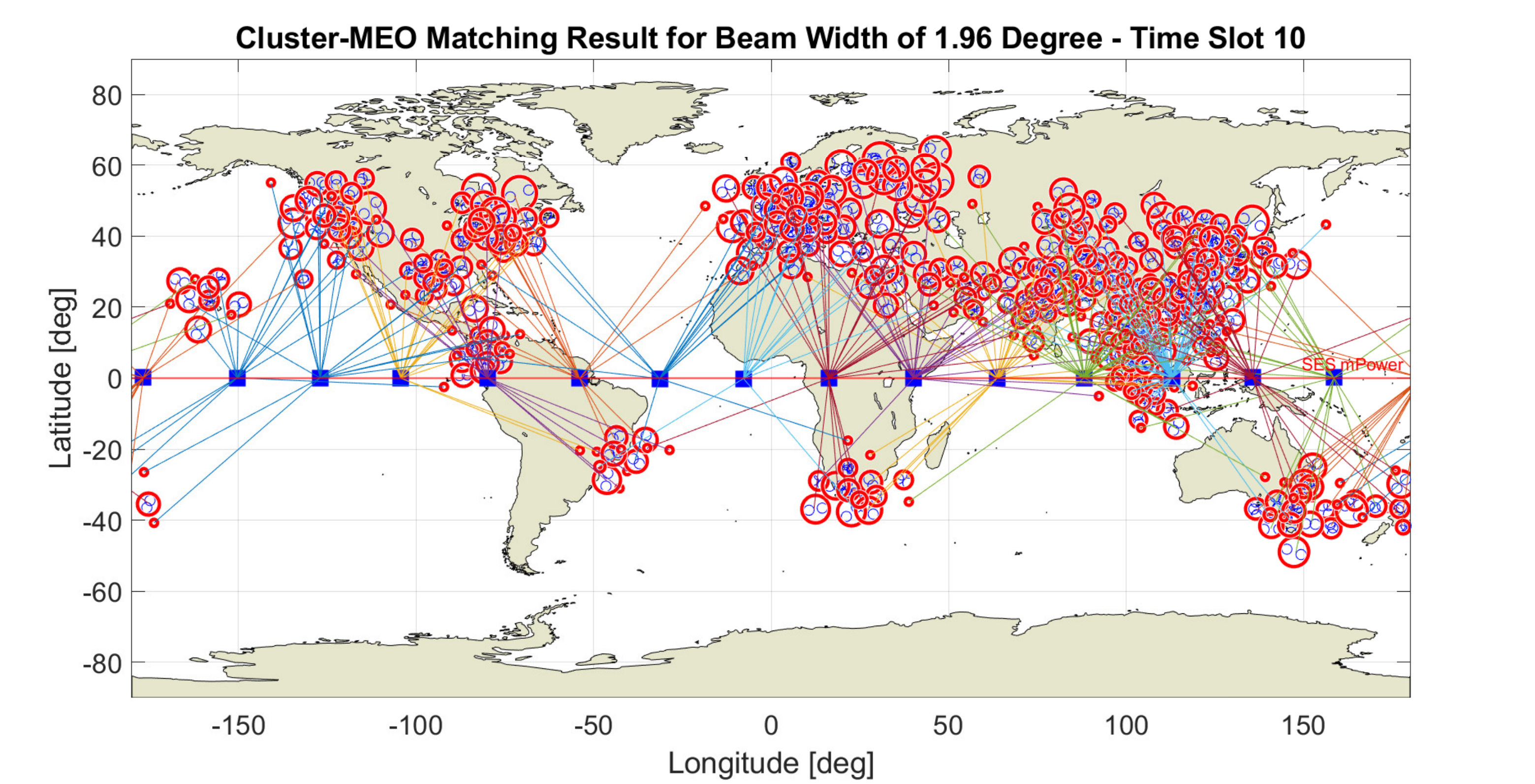}
	\caption{Matching result for angle threshold $1.96^{\circ}$ at TS $10$.}
	\label{Cluster-MEO-Matching196_TS10} \vspace{-3mm}
\end{figure}

\begin{figure}[!t]
	\centering
	\includegraphics[width=90mm]{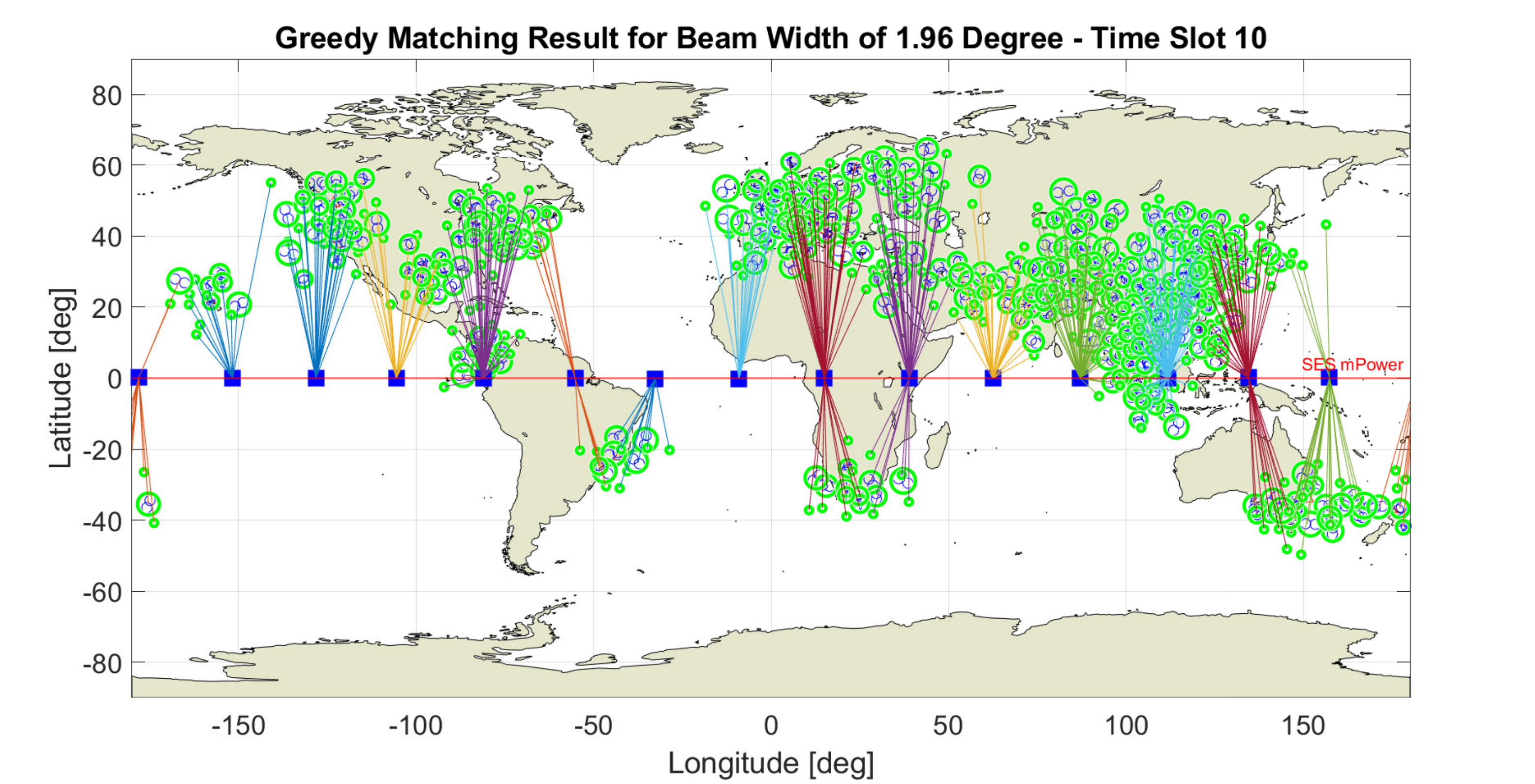}
	\caption{Greedy matching result for angle threshold $1.96^{\circ}$ at TS $10$.}
	\label{Cluster-MEO-Matching196_TS40} \vspace{-3mm}
\end{figure}

The results of Cluster-MEO matching obtained through the proposed and greedy algorithms for TS 10 are displayed in Figs.~\ref{Cluster-MEO-Matching196_TS10} and \ref{Cluster-MEO-Matching196_TS40}, respectively. The minimum and maximum number of clusters connected to an MEO, the ratios of served clusters, and the number of satisfied users are also summarized in Table~\ref{cluster_MEO_Mathcing_result}. These figures indicate that the proposed method provides a better dynamic beam placement solution compared to the greedy algorithm. Specifically, the nearest-MEO selection framework results in an uneven load distribution across all MEOs, with some MEOs serving a larger number of clusters than others. When an MEO becomes overloaded, we propose to step-by-step remove the cluster with the highest demand until it can accommodate all remaining clusters. Interestingly, the proposed approach is able to satisfy a greater number of users compared to the greedy algorithm.

\begin{figure}[!t]
	\centering
	\includegraphics[width=85mm]{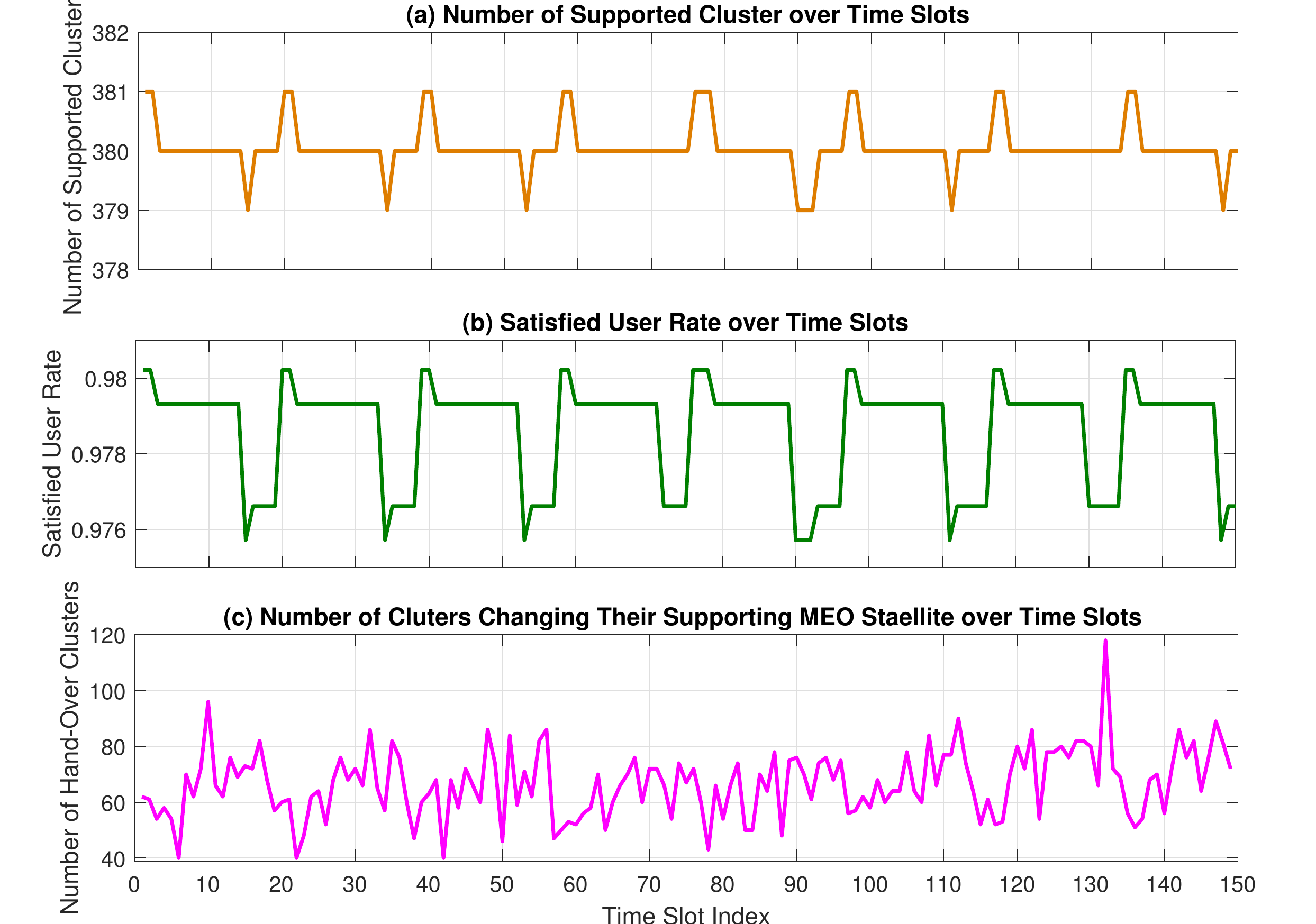}
	\caption{The results of the proposed algorithm over $150$ TSs.}
	\label{No_Supp_Clus_150TS} \vspace{-3mm}
\end{figure}

The results of the proposed algorithm during the observed time window are shown in Fig.~\ref{No_Supp_Clus_150TS}. In particular, this figure illustrates the number of supported clusters, the satisfied user rates, and the number of hand-over clusters. The satisfied user rate in each time slot is calculated as the ratio of the number of users whose traffic demand is fulfilled to the total number of users in the simulation. Interestingly, the number of supported clusters and the satisfied user rate both fluctuate periodically with an average cycle duration of approximately $16$ TSs, which is also the required time for an MEO satellite to reach the position of its nearest MEO. As seen in Fig.~\ref{No_Supp_Clus_150TS}-b, the high satisfactory rate, ranging from 0.976 to over 0.98, confirms the efficiency of the proposed algorithm. Additionally, Fig.~\ref{No_Supp_Clus_150TS}-c shows the number of clusters that change their supporting MEO after each TS, which varies irregularly over time.

\begin{figure}[!t]
	\centering
	\includegraphics[width=85mm]{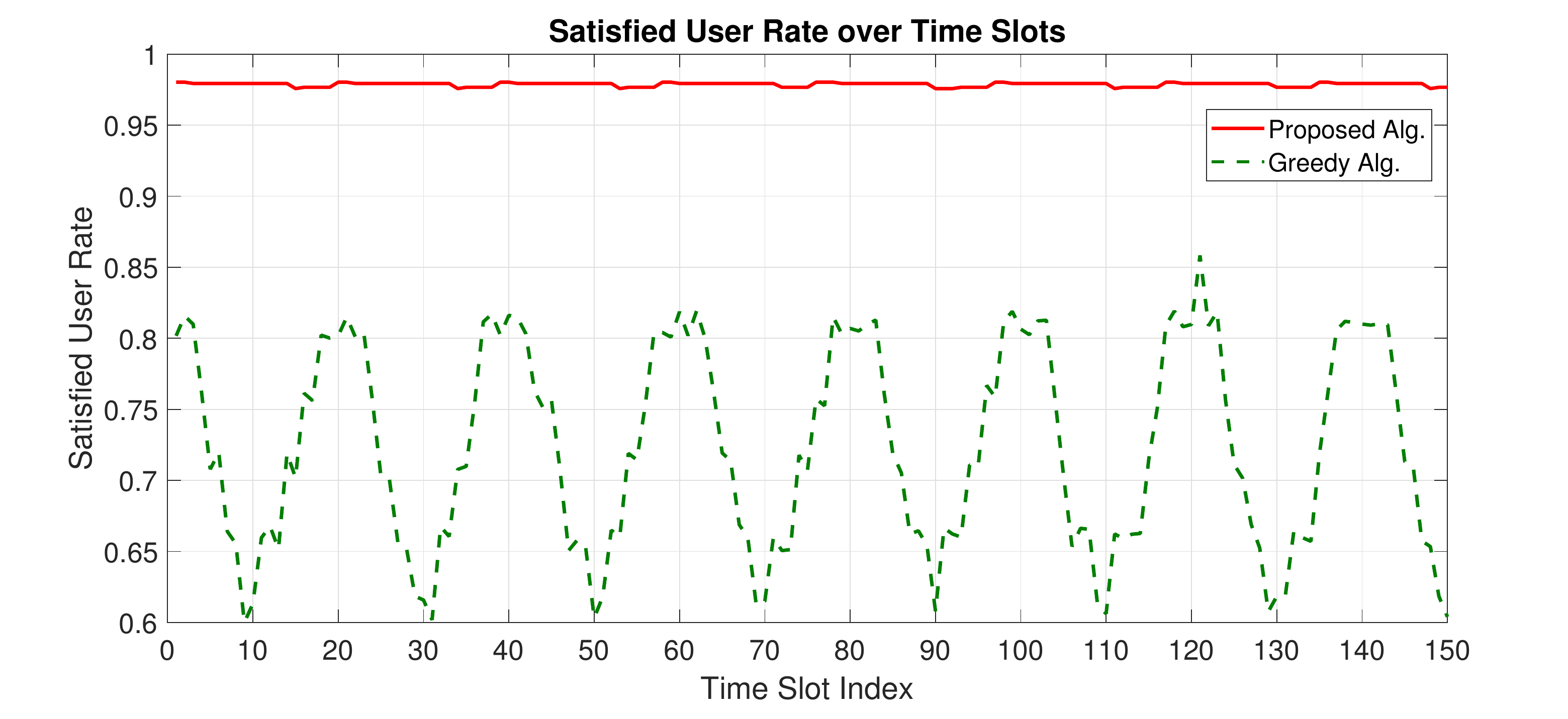}
	\caption{Satisfied user rate over $150$ TSs.}
	\label{TxPower_150TS} \vspace{-3mm}
\end{figure}

\begin{figure}[!t]
	\centering
	\includegraphics[width=85mm]{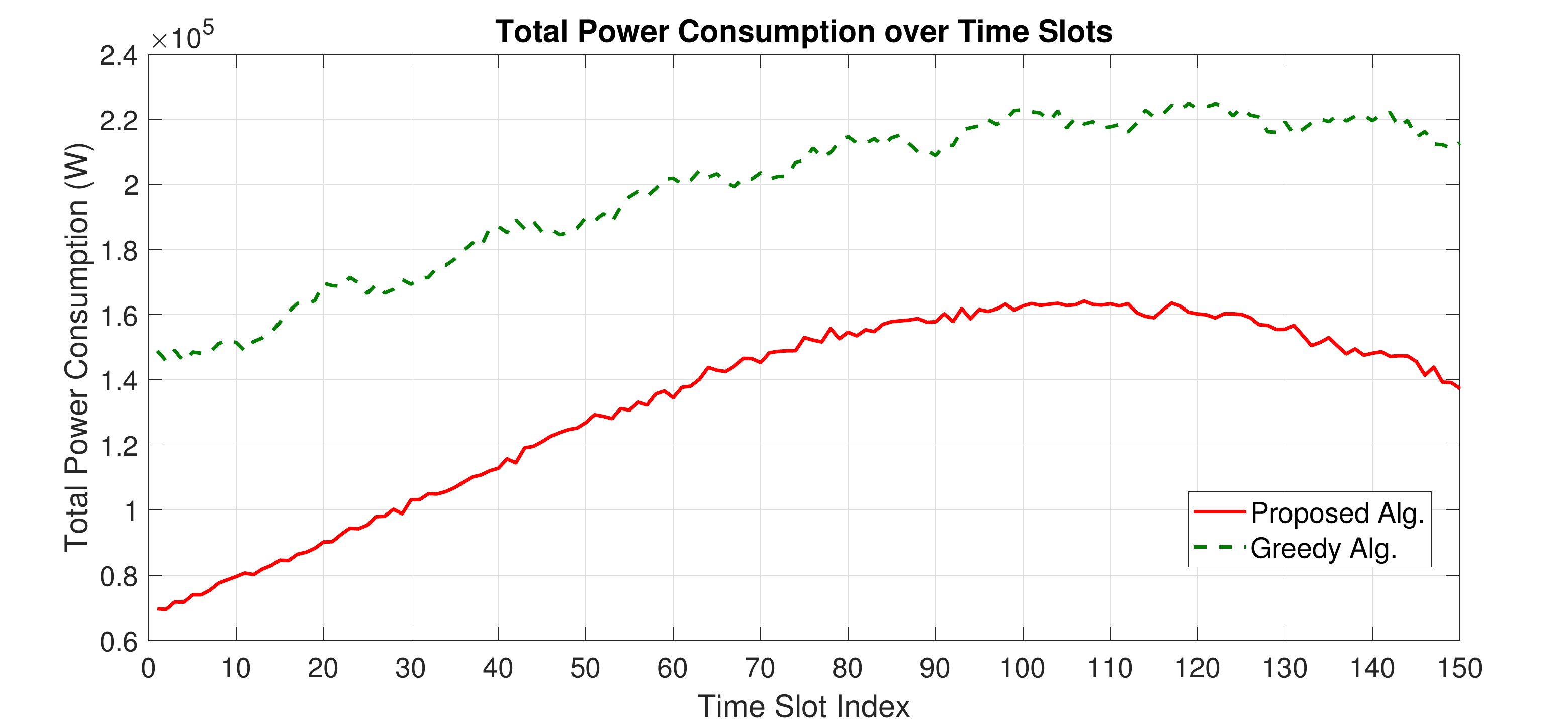}
	\caption{Total power consumption over $150$ TSs.}
	\label{BW_150TS} \vspace{-3mm}
\end{figure}
Finally, Figs.~\ref{TxPower_150TS} and \ref{BW_150TS} illustrate the satisfied user rate and the total power consumption corresponding to the proposed and greedy algorithms over $150$ TSs.
As can be observed, the proposed approach outperforms the greedy algorithm in terms of satisfying user demands and conserving energy. Our scheme is able to meet the needs of more than $97\%$ of the users, while the greedy algorithm only supports a maximum of $85\%$, but its uses over a third of our total power consumption. Fig.~\ref{BW_150TS} reveals a clear correlation between the power consumption and user demand over the time window, with higher power consumption observed during the higher-demand TSs.

%\vspace{-3mm}

\section{Conclusion}
In conclusion, this work presents a centralized power-minimization framework for the joint beam placement, power and bandwidth allocation design in MEO constellations to meet the QoS requirements of all users. The proposed three-stage solution approach addresses the challenging NP-hard problem and provides a practical solution for large-scale MEO constellations with a large number of global users. The results demonstrate that the proposed framework effectively reduces power consumption while meeting the QoS requirements of over $97\%$ of users. 

\section*{Acknowledgment}
This work is supported in parts by the Luxembourg National Research Fund (FNR) projects FlexSAT (C19 /IS/13696663), SmartSpace (C21/IS/16193290), MegaLEO (C20/IS/14767486), and the SES-Contract project MEOBRA.

\bibliographystyle{IEEEtran}
%\bibliography{ref_satnex,references}

\begin{thebibliography}{10}
\providecommand{\url}[1]{#1}
\csname url@samestyle\endcsname
\providecommand{\newblock}{\relax}
\providecommand{\bibinfo}[2]{#2}
\providecommand{\BIBentrySTDinterwordspacing}{\spaceskip=0pt\relax}
\providecommand{\BIBentryALTinterwordstretchfactor}{4}
\providecommand{\BIBentryALTinterwordspacing}{\spaceskip=\fontdimen2\font plus
\BIBentryALTinterwordstretchfactor\fontdimen3\font minus
  \fontdimen4\font\relax}
\providecommand{\BIBforeignlanguage}[2]{{%
\expandafter\ifx\csname l@#1\endcsname\relax
\typeout{** WARNING: IEEEtran.bst: No hyphenation pattern has been}%
\typeout{** loaded for the language `#1'. Using the pattern for}%
\typeout{** the default language instead.}%
\else
\language=\csname l@#1\endcsname
\fi
#2}}
\providecommand{\BIBdecl}{\relax}
\BIBdecl

\bibitem{Al-Hraishawi_CST2022}
H.~Al-Hraishawi, H.~Chougrani, S.~Kisseleff, E.~Lagunas, and S.~Chatzinotas,
  ``A survey on non-geostationary satellite systems: The communication
  perspective,'' \emph{IEEE Communications Surveys \& Tutorials}, pp. 1--1,
  2022.

\bibitem{Su2019}
Y.~Su, Y.~Liu, Y.~Zhou, J.~Yuan, H.~Cao, and J.~Shi, ``Broadband leo satellite
  communications: Architectures and key technologies,'' \emph{IEEE Wireless
  Communications}, vol.~26, no.~2, pp. 55--61, 2019.

\bibitem{Pachler_IJSCN2021}
N.~Pachler de~la Osa, M.~Guerster, I.~del Portillo~Barrios, E.~Crawley, and
  B.~Cameron, ``Static beam placement and frequency plan algorithms for leo
  constellations,'' \emph{International Journal of Satellite Communications and
  Networking}, vol.~39, no.~1, pp. 65--77, 2021.

\bibitem{Liu_GC2020}
B.~Liu, C.~Jiang, L.~Kuang, and J.~Lu, ``Joint user grouping and beamwidth
  optimization for satellite multicast with phased array antennas,'' in
  \emph{GLOBECOM}, 2020, pp. 1--6.

\bibitem{Phuc_MeditCon2022}
V.-P. Bui, T.~Van~Chien, E.~Lagunas, J.~Grotz, S.~Chatzinotas, and
  B.~Ottersten, ``Joint beam placement and load balancing optimization for
  non-geostationary satellite systems,'' in \emph{IEEE MeditCom}, 2022.

\bibitem{Xia_Access2019}
S.~Xia, Q.~Jiang, C.~Zou, and G.~Li, ``Beam coverage comparison of leo
  satellite systems based on user diversification,'' \emph{IEEE Access},
  vol.~7, pp. 181\,656--181\,667, 2019.

\bibitem{Hung_WSA23}
H.~Nguyen-Kha, V.~N. Ha, E.~Lagunas, S.~Chatzinotas, and J.~Grotz,
  ``Leo-to-user assignment and resource allocation for uplink transmit power
  minimization,'' in \emph{IEEE WSA \& SCC 2023}.

\bibitem{VuHa_TWC22}
L.~Chen, V.~N. Ha, E.~Lagunas, L.~Wu, S.~Chatzinotas, and B.~Ottersten, ``The
  next generation of beam hopping satellite systems: Dynamic beam illumination
  with selective precoding,'' \emph{IEEE Transactions on Wireless
  Communications}, 2022.

\bibitem{VuHaGC2022}
V.~N. Ha, T.~T. Nguyen, E.~Lagunas, J.~C. Merlano~Duncan, and S.~Chatzinotas,
  ``{GEO} payload power minimization: Joint precoding and beam hopping
  design,'' in \emph{GLOBECOM 2022 - 2022 IEEE Global Commun. Conf.}, 2022, pp.
  6445--6450.

\bibitem{3gpp2018study}
``{ Study on new radio (NR) to support nonterrestrial networks},'' 3GPP TR
  38.811 V1. 0.0, Stage1 (Release 15), 2018.

\bibitem{Hung_ICCWks23}
H.~Nguyen-Kha, V.~N. Ha, E.~Lagunas, S.~Chatzinotas, and J.~Grotz,
  ``'leo-to-user assignment and resource allocation for uplink transmit power
  minimization,'' in \emph{2023 IEEE ICC Workshops (GC Wkshps)}, 2023.

\bibitem{Angeletti_Access20}
P.~Angeletti and R.~De~Gaudenzi, ``{A Pragmatic Approach to Massive MIMO for
  Broadband Communication Satellites},'' \emph{IEEE Access}, vol.~8, pp.
  132\,212--132\,236, 2020.

\bibitem{VuHa_VTCFall22}
E.~Lagunas, V.~N. Ha, T.~V. Chien, S.~Andrenacci, N.~Mazzali, and
  S.~Chatzinotas, ``Multicast {MMSE}-based precoded satellite systems: User
  scheduling and equivalent channel impact,'' in \emph{2022 IEEE 96th Veh.
  Tech. Conf. (VTC2022-Fall)}, 2022, pp. 1--6.

\bibitem{VuHa_Access17}
V.~N. Ha and L.~B. Le, ``End-to-end network slicing in virtualized
  {OFDMA}-based cloud radio access networks,'' \emph{IEEE Access}, vol.~5, pp.
  18\,675--18\,691, 2017.

\bibitem{Sanjabi_TSP14}
M.~Sanjabi, M.~Razaviyayn, and Z.-Q. Luo, ``Optimal joint base station
  assignment and beamforming for heterogeneous networks,'' \emph{IEEE Trans.
  Signal Process.}, vol.~62, no.~8, pp. 1950--1961, 2014.

\bibitem{VuHa_TWC18}
V.~N. Ha, D.~H.~N. Nguyen, and J.-F. Frigon, ``Subchannel allocation and hybrid
  precoding in millimeter-wave {OFDMA} systems,'' \emph{IEEE Trans. Wireless
  Commun.}, vol.~17, no.~9, pp. 5900--5914, 2018.

\bibitem{VuHa_GCWks22}
V.~N. Ha and et~al., ``Joint linear precoding and {DFT} beamforming design for
  massive {MIMO} satellite communication,'' in \emph{2022 IEEE Globecom
  Workshops (GC Wkshps)}, 2022, pp. 1121--1126.

\bibitem{MaNing_JIoT2022}
N.~Ma, H.~Zhang, H.~Hu, and Y.~Qin, ``Escvad: An energy-saving routing protocol
  based on voronoi adaptive clustering for wireless sensor networks,''
  \emph{IEEE Internet of Things Journal}, vol.~9, no.~11, pp. 9071--9085, 2022.

\end{thebibliography}

%\begin{thebibliography}{1}
%\bibitem{3gpp2018study}
%3GPP TR 38.811 V1. 0.0, ``Study on new radio (NR) to support nonterrestrial networks,'' Stage1 (Release 15), 2018.
%\end{thebibliography}



%\vfill


\end{document}